# Position Sensorless and Adaptive Speed Design for Controlling Brushless DC Motor Drives

S. Geraee[a], M. Shafiei[b], A. R. Sahami[c], S. Alavi[d]

a: School of Electrical and Computer Engineering, University of Tabriz, Tabriz, Iran
b: School of Electrical Engineering and Computer Science, Queensland University of Technology, Brisbane, Australia
c: School of Electrical and Computer Engineering University of North Carolina at Charlotte Charlotte, NC
d: School of Electrical and Electronics Engineering, Shiraz University of Technology, Shiraz, Iran
shiva_co28@yahoo.com, m.shafiei@qut.edu.au, asahami@uncc.edu, saeidalavi@gmail.com

*Abstract*—This paper proposes a method for direct torque control of Brushless DC (BLDC) motors. Evaluating the trapezium of back-EMF is needed, and is done via a sliding mode observer employing just one measurement of stator current. The effect of the proposed estimation algorithm is reducing the impact of switching noise and consequently eliminating the required filter. Furthermore, to overcome the uncertainties related to BLDC motors, Recursive Least Square (RLS) is regarded as a real-time estimator of inertia and viscous damping coefficients of the BLDC motor. By substituting the estimated load torque in mechanical dynamic equations, the rotor speed can be calculated. Also, to increase the robustness and decrease the rise time of the system, Modified Model Reference Adaptive System (MMRAS) is applied in order to design a new speed controller. Simulation results confirm the validity of this recommended method.

## I. INTRODUCTION

Synchronous motors are gaining much attention compared with induction motors [1-6]. Brushless DC (BLDC) motors are classified as synchronous AC electric motors, which play a significant role in contemporary applications, such as: industry, transportation, visual and sound equipment in medical applications, with a large variety of power ranges. Preferable speed-torque features, great dynamic response, long life cycle, noiseless action, and broad speed dimensions are the advantages of BLDC motors compared with brushed DC and induction motors.

One of the most ubiquitous control drives of BLDC motors is Direct Torque Control (DTC), which has been presented in [7,8]. It is worth mentioning that the conventional DTC method, which has been presented for Permanent Magnet Synchronous Motors (PMSMs), cannot be employed for BLDC motors. The DTC in BLDC motors is simple with high dynamic torque responses and low sensitivity to parameter variations.

BLDC motors operate in either a constant current region or a constant power region. The raising in phase current and consequently output torque can devastate the control drives algorithm in a constant torque region. However, in a constant torque region, the BLDC motor's back-EMF is less than the DV voltage of the inverter, which can prevent phase current rises. Therefore, in this study the simulations are based on the constant torque operating region.

Commutation has to be done at exact rotor positions for efficient performance. Hence, six discrete rotor positions should be considered to feed BLDC motors with rectangular phase currents. Although Hall sensors are the best type of sensors for BLDC motors due to soft starting, it has several drawbacks, such as: cost, sensitivity to the temperature, and unreliable performance in industrial areas. To overcome these problems diverse approaches have been addressed in articles [9-16]. For example, the proposed methods for folding trace technique of Magnet Resistors as well as shielding for Hall effect sensing element [17] could enhance bandwidth for advanced current controlling schemes. Other approaches include mechanisms based on integration and third harmonics of back-EMF, flux and terminal voltage calculation, and discovery of freewheeling diodes.

For BLDC motors, sensorless operation is one of the effective methods in speed-dependent back-EMF. However, this method has its own drawbacks, such as low performance in low speed BLDC motors, and position error at the estimated commutation point.

Extended Kalman Filter (EKF) and sliding mode algorithms are presented in [18-23]. The drawbacks of EKF are: noise influence, complicated configuration, and computation implementation adversity. In contrast, sliding mode observers' advantages, such as: being appropriate for piecemeal designing, being suitable for non-smooth systems, having firmness against parameters variation, converging in the shortest possible time, and reducing the observation error, has made it an efficient popular method [24]. Therefore, sliding mode observer is employed in the current study. The Recursive Least Square (RLS) is used to estimate the rotor inertia and the viscous damping coefficient [19] so that, the load torque can be estimated. Next, the rotor speed is calculated based on the estimated load torque using mechanical dynamic equations. Finally, measured stator currents, estimated back-EMF, and rotor speed are used to calculate electromagnetic torque.

Model Reference Adaptive System (MRAS) is a closed loop control system and its parameters can be adjusted to decrease the error between actual models and reference models [25, 26]. In this paper, MIT role is employed to design the MRAS. The MRAS is used instead of a PI controller, in order to gain better speed and torque responses.

## II. BLDC MOTOR

The sharp distinction between brushed DC motors and BLDC motors is the procedure for commutation. In common DC motors, brushes and mechanical commutators are responsible for commutation, while in BLDC motors, an electronic switching converter performs the commutation. BLDC motors as AC synchronous motors are designed such that permanent magnets are located on the rotor surface, and the stator is wire-wounded. The exciting process in BLDC motors is because of the permanent magnets as a constant flux source which is called rotor flux. The rotor is the rotating part of a motor and it needs a rotating field, which is provided by stator windings. Hence, by employing the right switching pattern, the perfect rotating field can be supplied. The ideal waveform of back-EMF and currents in BLDC motors with three-phase and 120° conductivity is presented in Fig. 1 [11-13].

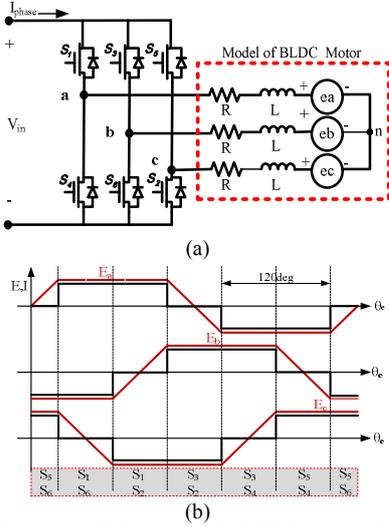

(a)

(b)

Fig. 1. (a) Structure of drive system in BLDC motor , (b) back EMF arrangement creation of reference current

The equations of BLDC motors can be depicted as follows:

$$\begin{bmatrix} V_a \\ V_b \\ V_c \end{bmatrix} = \begin{bmatrix} R_a & 0 & 0 \\ 0 & R_b & 0 \\ 0 & 0 & R_c \end{bmatrix} \begin{bmatrix} i_a \\ i_b \\ i_c \end{bmatrix} + \begin{bmatrix} L & 0 & 0 \\ 0 & L & 0 \\ 0 & 0 & L \end{bmatrix} \times \frac{d}{dt} \begin{bmatrix} i_a \\ i_b \\ i_c \end{bmatrix} + \begin{bmatrix} e_a \\ e_b \\ e_c \end{bmatrix} \quad (1)$$

In order, $u_a$, $u_b$, and $u_c$ are the voltages of each phases, while $i_a$, $i_b$, and $i_c$ are their respective currents. The back-EMF waveforms of phases are named $e_a$, $e_b$ and $e_c$. R and L are symbols of the resistance and the inductance per phase, and $M$ shows the mutual inductance per phase. The electromagnetic torque is achieved by:

$$T_e = \frac{1}{\omega_r}(e_a i_a + e_b i_b + e_c i_c) \quad (2)$$

where $\omega_r$ presents the motor speed. Furthermore, the dynamic motion is obtained by using the following equation:

$$\frac{d}{dt}\omega_r = \frac{1}{J}(T_e + T_L + B\omega_r) \quad (3)$$

where $B$ is the damping coefficient, J is the motor shaft and load inertias, $T_L$ is the mechanical torque.

## III. DTC IN BLDC MOTORS DRIVE

DTC stands for Direct Torque Control. The history of this control method goes back to 1986, when Takahashi and Noguchi presented this method. The strength of DTC is its ability to control electromagnetic torque and flux linkage simultaneously. Therefore, DTC is the appropriate choice when exact torque control is necessary for BLDC motors. The look-up table in [7] shows the six voltage space vectors which are used to implement DTC. A suitable voltage space vector should be chosen for employing DTC in BLDC motors. Since the torque of BLDC motors should be limited to the hysteresis band by turning off/on transistors, the switching losses are not significant. As noted earlier, the electromagnetic torque quantity plays an important role in the DTC algorithm, especially in a constant torque region. Thus, the final form of the electromagnetic equation in the stationary reference frame is eventuated as [7, 8, 13, 14]:

$$T_e = \frac{3}{2}\frac{P}{2}\frac{1}{\omega_e}(e_{sq}i_{sq} + e_{sd}i_{sd}) = \frac{3}{2}\frac{P}{2}[k_q(\theta_e)i_{sq} + k_d(\theta_e)i_{sd}] \quad (4)$$

where $\theta_e$ and $\omega_e$ are the electrical rotor position and speed, $k_q(\theta_e)$, $k_d(\theta_e)$, $e_{sq}$, $e_{sd}$, $i_{sq}$ and $i_{sd}$ are back-EMF coefficients, back-EMFs, and stator currents in the stationary reference frame (dq-axes) respectively.

The general configuration of DTC, which is performed in BLDC motors, is demonstrated in Fig.2, which shows several parts such as: a speed controller, an electromagnetic torque estimator, a comparator block for torques, and a switching table. Based on the error between the actual and the reference speed, the reference torque can be calculated. Switching commands, which are instructed to inverters, are based on the hysteresis controller results.

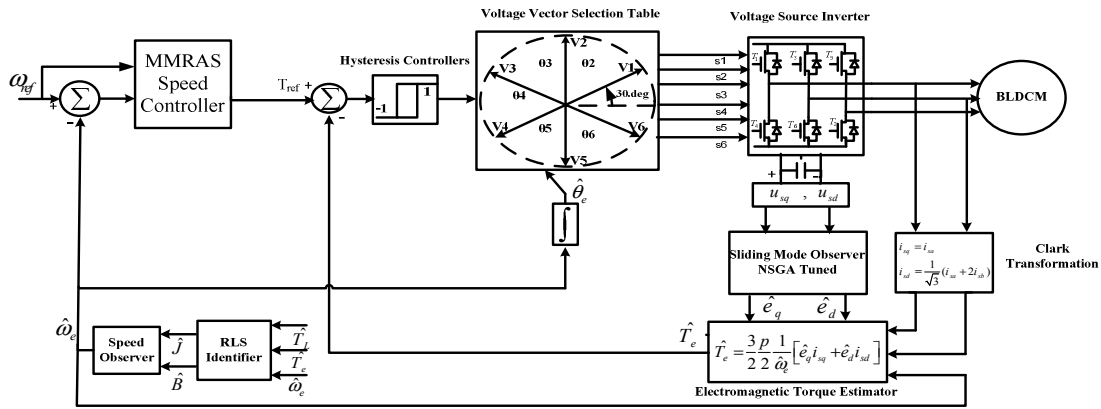

Fig. 2. The general configuration of in BLDC motors drive

TABLE I
SWITCHING MODES TABLE FOR DTC, INCREASING TORQUE (TI), DECREASING TORQUE (TD) UNCHANGES FLUX (F)

| $F_{st}$ | $T_{st}$ | Sectors | | | | | |
|---|---|---|---|---|---|---|---|
| | | Θ1 | Θ2 | Θ3 | Θ4 | Θ5 | Θ6 |
| F | TI | V2(001001) | V3(011000) | V4(010010) | V5(000110) | V6(100100) | V1(100001) |
| | TD | V5(000110) | V6(100100) | V1(100001) | V2(001001) | V3(011000) | V4(010010) |

Hysteresis control is responsible for correlating real quantities with the commanded torque. To have an appropriate performance, a template for switching is necessary based on flux and torque error, $F_{st}$, $T_{st}$ respectively. In DTC, both torque and flux are regularly controlled. However, the flux control is not considered in designing the DTC for BLDC motors. Therefore, in switching look-up table, $F_{st}$ is set as zero permanently. Consequently, the only effective factor is torque error, which is the discrepancy between the actual and the commanded electromagnetic torque quantity. The number assigned to $T_{st}$ (the torque error) is based on the hysteresis controller output. For states which the real torque takes less extent than the reference torque in the hysteresis band, $T_{st}$ is named "TI" (Increasing Torque). In other cases, it is specified as "TD" (Decreasing Torque). The proper functioning of DTC is expected if the switching table is applied correctly.

## IV. MMRAS SPEED CONTROLLER

In this paper MRAS, as one of the most noteworthy adaptive control approaches, is proposed [11]. Fig. 3 depicts the general formation of the system. Three parts are combined together to shape the feedback of the system in the following order: processor, controller, and adaption loop. The compared result of the desired model output ($y_{ref}$) and actual system output ($y$) determines the parameters of the controller. The MIT rule is the name of the method which is offered to process error ($e$) in terms of finding controller parameters [27-29]. Fig. 4 illustrates that the plant of the system is eventuated from speed control, and it can be equivalent to a second order system as presented in equation (5).

$$G_p(s) = \frac{y(s)}{u'(s)} = \frac{a}{s^2 + bs + c} \quad (5)$$

where $a = R/L+B/J$, $b = RB/LJ$, $c = k\varphi fa/LJ$ and $s$ is Laplace operator.

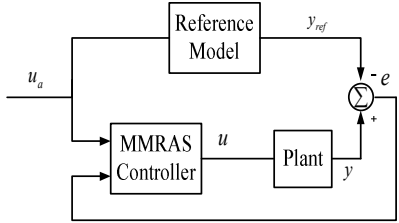

Fig. 3. Modified model reference adaptive system

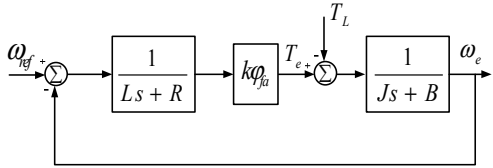

Fig. 4. Block diagram of close loop speed controller

The following equation represents the reference model in the same way as equation (5):

$$G_{ref}(s) = \frac{y_{ref}(s)}{u(s)} = \frac{a'}{s^2 + b's + c'} \quad (6)$$

where $y_{ref}$ and $u$ are the output of the model reference and the reference signal respectively. Parameters $a'$, $b'$ and $c'$ are constants that should be adjusted to obtain the desired model reference.

The control law of the MMRAS control is:

$$u' = \theta_1 u - \theta_2 e \quad (7)$$

where $e$ relies on $y - y_{ref}$, and $\theta_1$ and $\theta_2$ are the adjustable parameters. Substituting (7) in (5), the relation between $y$ and $y_{ref}$ is achieved as [21]:

$$y = \frac{a\theta_1}{s^2 + bs + (c + a\theta_2)} u + \frac{a\theta_2}{s^2 + bs + (c + a\theta_2)} y_{ref} \quad (8)$$

The MIT rule is applied to minimize the loss function:

$$j(\theta) = (\frac{1}{2})e^2$$

In the loss function, $e$ is a customizable controller parameter. To minimize the loss function, the following equation is employed:

$$\frac{d\theta}{dt} = -\gamma \frac{\delta j}{\delta \theta} = -\gamma e \frac{\delta e}{\delta \theta} \quad (9)$$

Adaption gain is the name of $\gamma$. [28]. When calculating sensitivity, which is derivative of controller parameters, the following equations are achieved:

$$\frac{\delta e}{\delta \theta_1} = \left(\frac{a}{s^2 + bs + (c + a\theta_2)}\right) u \quad (10)$$

$$\frac{\delta e}{\delta \theta_2} = \left(\frac{-a\theta_1}{(s^2 + bs + (c + a\theta_2))^2}\right) u$$

$$= \left(\frac{a}{s^2 + bs + (c + a\theta_2)}\right) y_{ref} \quad (11)$$

$$= \left(\frac{-a^2 \theta_2}{(s^2 + bs + (c + a\theta_2))^2}\right) y_{ref}$$

The amount of the output control is low in the steady state. So, the $a\theta_2$ is insignificant and equations (10) and (11) can be rewritten as:

$$\frac{\delta e}{\delta \theta_1} = \left(\frac{a}{s^2 + bs + c}\right) u \quad (12)$$

$$\frac{\delta e}{\delta \theta_2} = -\left(\frac{a}{s^2 + bs + c}\right) e \quad (13)$$

Considering (12) and (13), the adaption laws are given as:

$$\frac{d\theta_1}{dt} = \gamma \left(\frac{a}{s^2 + bs + c}\right) u \quad (14)$$

$$\frac{d\theta_2}{dt} = -\gamma \left(\frac{a}{s^2 + bs + c}\right) e \quad (15)$$

## V. OBSERVER TECHNIQUE

Sliding mode observer is known as a favorable observer of non-linear category observers, and is discussed in detail in [30]. The quality of this method returns to its capability of damping the estimator error quickly. Subsequently, estimates of the observer reconcile with the real outcome of the system. The following equations can be obtained for the BLDC motor (16):

$$\frac{di_{sq}}{dt} = -\frac{R}{L}i_{sq} - \frac{1}{L}e_{sq} + \frac{1}{L}u_{sq}$$
$$\frac{di_{sd}}{dt} = -\frac{R}{L}i_{sd} - \frac{1}{L}e_{sd} + \frac{1}{L}u_{sd}$$
$$\frac{de_{sq}}{dt} = 0$$
$$\frac{de_{sd}}{dt} = 0 \quad (16)$$

In situations where the electrical and the mechanical time constants are greater than the sampling period, variations of the back-EMF can be negligible within each sampling period [20]. Hence,

$$de_{sq}/dt = 0, \quad de_{sd}/dt = 0$$

Defining $x_1 = i_{sq}$, $x_2 = i_{sd}$, $x_3 = e_{sq}$ and $x_4 = e_{sd}$, the state-space equations of the system given in (6) can be written as follows:

$$\dot{x}_1 = -\alpha_1 x_1 + \alpha_2(-x_3 + u_1)$$
$$\dot{x}_2 = -\alpha_1 x_2 + \alpha_2(-x_4 + u_2)$$
$$\dot{x}_3 = 0$$
$$\dot{x}_4 = 0 \quad (17)$$
$$y_1 = x_1$$
$$y_2 = x_2$$

In (17), $u_1 = u_{sq}$ and $u_2 = u_{sd}$ are the input variables, $y_1$ and $y_2$ represent the output variables, and $\alpha_1 = R/L$ and $\alpha_2 = 1/L$ are the system parameters.

The sliding mode observer is proposed as:

$$\dot{\hat{x}}_1 = -\alpha_1 x_1 + \alpha_2(-\hat{x}_3 + u_1) + k_{s1} sat(x_1 - \hat{x}_1)$$
$$\dot{\hat{x}}_2 = -\alpha_1 x_2 + \alpha_2(-\hat{x}_4 + u_2) + k_{s2} sat(x_2 - \hat{x}_2)$$
$$\dot{\hat{x}}_3 = k_{s3} sat(x_1 - \hat{x}_1) \quad (18)$$
$$\dot{\hat{x}}_4 = k_{s4} sat(x_2 - \hat{x}_2)$$

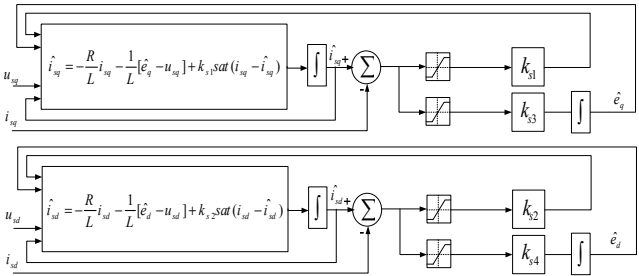

Fig. 5. Block diagram of sliding mode observer

The gains of the observer must be determined so that the inequality $S \cdot \dot{S} < 0$ is satisfied. To achieve the optimal gains of the observer, the NSGA-II is employed. Fig. 5 demonstrates the configuration of the sliding mode observer. There are several factors, such as unsteadiness, dynamics of the motor, and environmental changes that prevent us from considering parameters B and J as constants. To diminish the repercussions of these fluctuations, the Recursive Least Square (RLS) technique is used. Equation (22) shows the motion control:

$$y(t) = \varphi^T(t).\theta(t) \quad (22)$$

where $y = (d/dt)\omega_r$, $\varphi(t) = \omega_r(T_e - T_L)$ and $\theta = \begin{bmatrix} \frac{B}{J} & \frac{1}{J} \end{bmatrix}$.

The RLS rules are [18, 22]:

$$\hat{\theta}(t) = \hat{\theta}(t-1) + K(t)(y(t) - \varphi^T(t)\hat{\theta}(t-1)) \quad (23)$$
$$K(t) = P(t-1)\varphi(t) =$$
$$P(t-1)\varphi(t)(\lambda I + \varphi^T(t)P(t-1)\varphi(t))^{-1} \quad (24)$$
$$P(t) = \frac{(I - K(t)\varphi^T(t))P(t-1)}{\lambda} \quad (25)$$

where $\lambda$ is named the forgetting factor. Hence, $\hat{B}$ and $\hat{J}$ are estimated by (23-25). It can be seen in Fig. 6, that $\hat{B}$ and $\hat{J}$ are parameters which are used in load torque estimation. By determining load torque, rotor speed in (3) will be specified. Another parameter is weighting factor $K$, which is multiplied in estimated load torque to recognize a robust system.

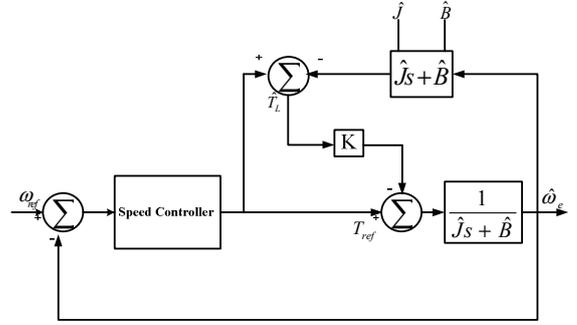

Fig. 6. Speed observer block diagram

Now all necessary parameters are identified to calculate torque as follows:

$$\hat{T}_e = \frac{3}{2}\frac{P}{2}\frac{1}{\hat{\omega}_e}(\hat{e}_q i_{sq} + \hat{e}_d i_{sd}) \quad (26)$$

And for electrical position of the rotor, the following equation is used:

$$\hat{\omega}_e = \frac{d}{dt}\hat{\theta}_e \quad (27)$$

## VI. SIMULATION RESULTS

Table 2 shows the parameters of the motor used in simulation.

TABLE 2
THE HANDLED BLDC MOTOR PARAMETER IN SIMULATION

| Parameter | Value | Unit |
|---|---|---|
| Number of Poles | 2 | [pole] |
| DC Link Voltage | 300 | [V] |
| Rated Speed | 1500 | [rpm] |
| Phase Resistance | 0.4 | [Ω] |
| Phase Inductance | 13 | [mH] |
| Load Torque | 3 | [N.m] |
| Moment of Inertia | 0.004 | [kg.m$^2$] |
| Torque Constant | 0.4 | [V/(rad/sec)] |
| Damping Constant | 0.002 | [N.m/(rad/sec)] |

The real and estimated back-EMFs in d-q frame are shown in Figs 8 and 9 in order. As can be observed, the estimated back-EMF is in accordance with the actual one. Hence, it proves the superior performance of the sliding mode observer rather than employing look-up table and position sensors.

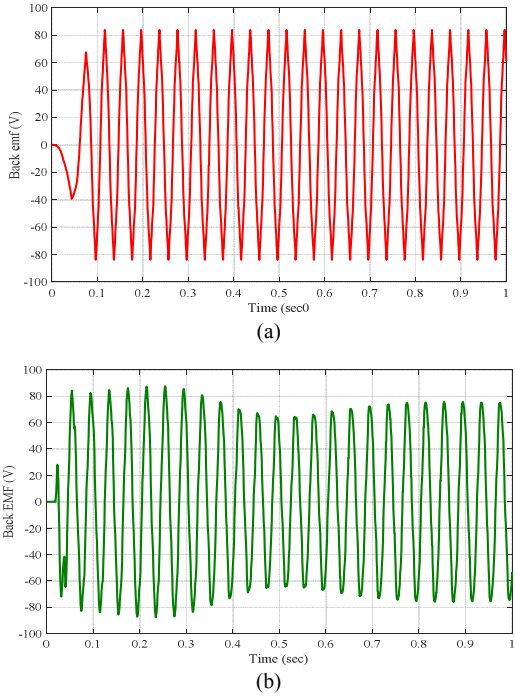

Figure. 8. Back-EMF ($e_q$), (a) Real value, (b) Estimated value

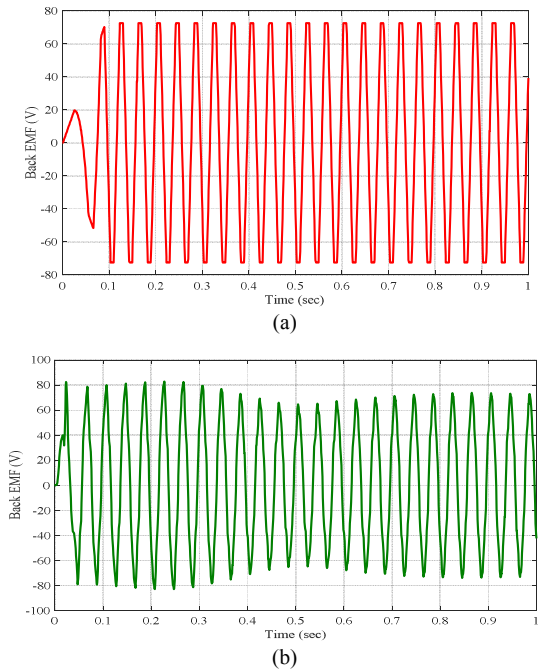

Fig. 9. Back-EMF ($e_d$), (a) Real value, (b) Estimated value

Figs. 10(a) and 10(b) depict the identified parameters $\hat{B}$ and $\hat{J}$ which confirm the efficiency of the RLS method.

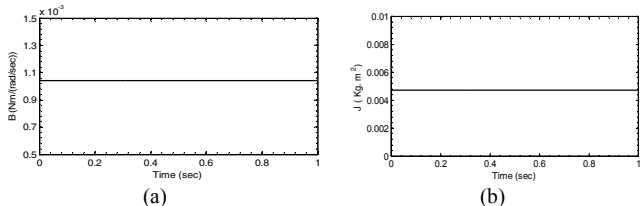

Fig. 10. Identified parameters, (a) Viscous damping coefficient, (b) moment of inertia

The estimated load torque is illustrated in Fig. 11 (a), and the estimated electromagnetic torque is presented in Fig. 11(b).

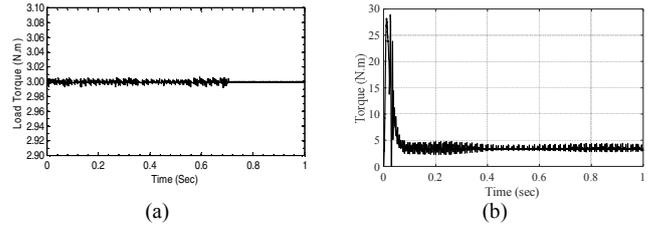

Fig. 11. (a) Estimated load torque, (b) Estimated electromagnetic torque in steady state mode

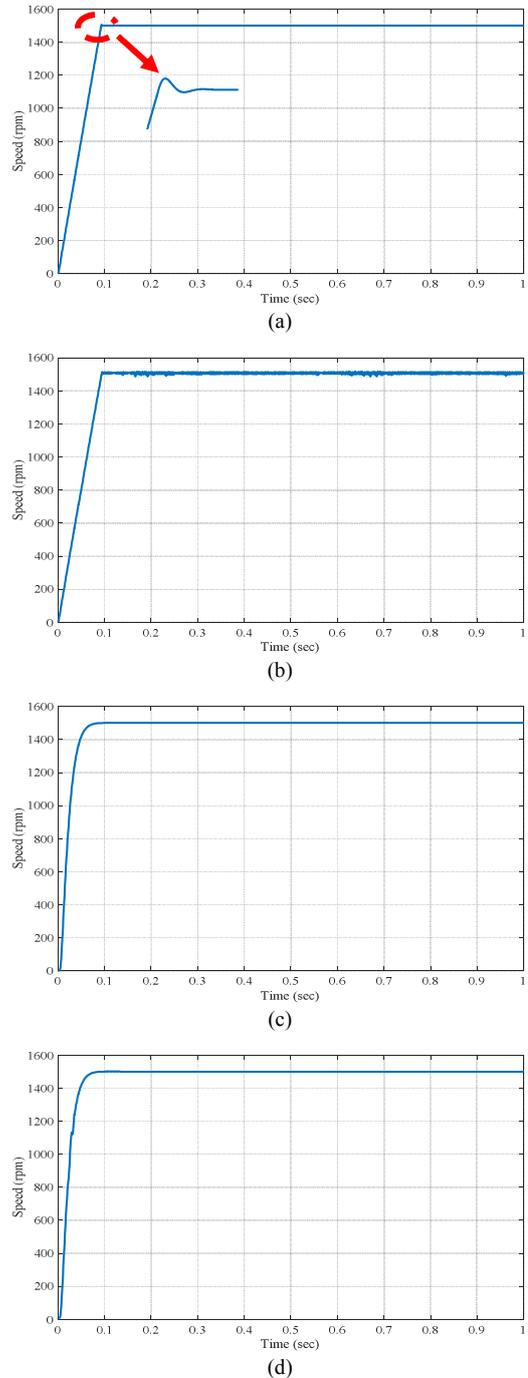

Fig. 12. Rotor speed, (a) Real value with PI Controller, (b) Estimated value with PI Controller, (c) Real value with MMRAS, (d) Estimated value with MMRAS

Based on the estimated and measured parameters, the PI and the MRAS speed controllers are compared in Fig. 12. The overshoot of the PI speed controller is damped completely by employing the MRAS controller.

VII. CONCLUSION

This paper has proposed an innovative plan to control brushless DC motor drives with no position sensor. In this plan, DTC also has been used due to its advantages. Another algorithm applied here is the RLS, which is used to assess some parameters of the speed observer in order to calculate the rotor speed. To satisfy the need for back-EMF estimation, a sliding mode observer is utilized. The proposed scheme can estimate the speed from standstill to the steady state. Simulation results have proven the success of the proposed approach.

REFERENCES

[1] A. Negahdari, V. M. Sundaram and H. A. Toliyat, "An analytical approach for determining harmonic cusps and torque dips in line start synchronous reluctance motors," 2016 IEEE Energy Conversion Congress and Exposition (ECCE), Milwaukee, WI, 2016, pp. 1-6.

[2] A. Negahdari and H. A. Toliyat, "Studying crawling effect in Line-Start Synchronous Reluctance Motors (LS-SynRM)," *2016 IEEE 25th International Symposium on Industrial Electronics (ISIE)*, Santa Clara, CA, 2016, pp. 210-215.

[3] Mozaffari Niapour SA-KH, Danyali S, Sharifian MBB, Feyzi MR. Brushless DC motor drives supplied by PV power system based on Z-source inverter and FLIC MPPT. Energy Convers Manage 2011;52(8–9):3043–59.

[4] S.A.K.H. Mozaffari Niapour, M. Tabarraie, and M. R. Feyzi "A new speed-sensorless robust control strategy for high-performance brushless DC motor drives with reduced torque ripple," Elsevier, Control Engineering Practice, vol. 64, pp. 42–54, Mar. 2014.

[5] S.A.KH. Mozaffari Niapour, M. Tabarraie, and M. R. Feyzi, "Design and analysis of speed-sensorless robust stochastic L∞-induced observer for high-performance brushless DC motor drives with diminished torque ripple," Elsevier, Energy Conversion and Management, vol. 64, pp. 482–98, Dec. 2012.

[6] M. R. Feyzi, S. A. K. Mozaffari Niapour, S. Danyali and M. Shafiei, "Supplying a Brushless DC Motor by z-source PV power inverter with FLC-IC MPPT by DTC drive," 2010 International *Conference on Electrical Machines and Systems*, Incheon, 2010, pp. 694-699.

[7] S.B. Ozturk, W.C Alexander, H.A. Toliyat. Direct Torque Control of Four-Switch Brushless DC Motor with Non sinusoidal Back-EMF. IEEE Transactions on Power Electronics 2010; 25(2): 263-271.

[8] Parhizkar, N., Shafiei, M., & Kouhshahi, M. B. (2011, August). Direct torque control of brushless DC motor drives with reduced starting current using fuzzy logic controller. In Uncertainty Reasoning and Knowledge Engineering (URKE), 2011 International Conference on (Vol. 1, pp. 129-132). IEEE.

[9] P. Damodharan, K. Vasudevan. Sensorless Brushless DC Motor Drive Based on the Zero-Crossing Detection of Back Electromotive Force (EMF) From the Line Voltage Difference. IEEE Transactions on Energy Conversion 2010; 25(3): 1-8.

[10] Yen-Shin Lai, Yong-Kai Lin. Novel Back-EMF Detection Technique of Brushless DC Motor Drives for Wide Range Control Without Using Current And Position Sensors. IEEE Transactions on Power Electronics 2010; 23(2): 934-940.

[11] Cheng-Hu, Chen Ming, Yang Cheng. A New Cost Effective Sensorless Commutation Method for Brushless DC Motors without Phase Shift Circuit and Neural Voltage. IEEE Transactions on Power Electronics 2007; 22(2): 644-653.

[12] Niapour, S. K. M., Garjan, G. S., Shafiei, M., Feyzi, M. R., Danyali, S., & Kouhshahi, M. B. (2014). Review of Permanent-Magnet Brushless DC Motor Basic Drives Based on Analysis and Simulation Study. International Review of Electrical Engineering (IREE), 9(5), 930-957.

[13] S. Ogasawara, H. Akagi. An approach to position sensorless drive for brushless DC motors. IEEE Transactions on Industry Application 1991; 27(5): 928-933.

[14] Sharifian, M. B. B., Shafiei, M., Sadeghi, M. S., & Golestaneh, F. (2011, August). Direct torque control of brushless DC motor drives based on ANFIS controller with fuzzy supervisory learning. In Electrical Machines and Systems (ICEMS), 2011 International Conference on (pp. 1-6). IEEE.

[15] M. Faeq, D. Ishak. A new scheme sensorless control of BLDC motor using software PLL and third harmonic back-EMF. IEEE Symposium on Industrial Electronics & Applications 2009; 861-865.

[16] Han. S-G, Renfrew. A.C. Sensorless torque vector speed control of a brushless DC motor. IET International Conference on Power Electronics 2002; 510-514.

[17] M. Biglarbegian, S. J. Nibir, H. Jafarian, B. Parkhideh, "Development of current measurement techniques for high frequency power converters", Intelec, October 2016

[18] H. Fakham, M. Djemai, K. Busawon. Design and practical implementation of a back-EMF sliding-mode observer for a brushless dc motor. IET Journal on Electric Power Application 2008; 2(6): 353-361.

[19] Shafiei, M., Kouhshahi, M. B., Sharifian, M. B. B., & Feyzi, M. R. (2011). Position sensorless for controlling brushless DC motor drives based on sliding mode and RLS estimators using NSGA-II algorithm optimization. system, 1, 2.

[20] Yong Liu, Zi Qiang Zhu, Howe D. Instantaneous Torque Estimation in Sensorless Direct-Torque-Controlled Brushless DC Motors. IEEE Transactions on Industry Applications 2006, 42(5): 1275-1283.

[21] Chen. Z, Tomita M., Doki S, Okuma S. New adaptive sliding observers for position- and velocity-sensorless controls of brushless DC motors. IEEE Transactions on Industrial Electronics 2000; 47(3): 582-591.

[22] Terzic B, Jadric M. Design and implementation of the extended Kalman filter for the speed and rotor position estimation of brushless DC motor. IEEE Transactions on Industrial Electronics 2001; 48(6): 1065-1073.

[23] Mingyao Lin, Weigang Gu, Wei Zhang, Qiang Li. Brushless DC motor sliding mode control with Kalman Filter. IEEE International Conference on Industrial Technology 2008; 1-6.

[24] Feyzi, M. R., Shafiei, M., Kouhshahi, M. B., & Niapour, S. K. M. (2011, February). Position sensorless direct torque control of Brushless DC motor drives based on sliding mode observer using NSGA-II Algorithm optimization. In Power Electronics, Drive Systems and Technologies Conference (PEDSTC), 2011 2nd (pp. 151-156). IEEE.

[25] H. M. Bagherpoor and F. R. Salmasi, "Robust model reference adaptive output feedback tracking for uncertain linear systems with actuator fault based on reinforced dead-zone modification," *ISA transactions,* vol. 57, pp. 51-56, 2015.

[26] Hsiu-Ping Wang, Yen-Tsan Liu. Integrated design of speed-sensorless and adaptive speed controller for a brushless DC motor. IEEE Transactions on Power Electronics 2006; 21(2): 518.

[27] K. J. Astrom and B. Wittenmark. Adaptive control. Addison-Wesley1989; 2d ed.

[28] Pirabakaran, K., Becerra, V.M. Automatic Tuning of PID Controllers Using Model Reference Adaptive Control Techniques. Industrial Electronics society 2001 IEDON'01 The 27th Annual Conference of the IEEE; 1: 736 – 740.

[29] K. J. Astrom and B. Wittenmark. Adaptive Control. 2nd ed. Reading MA Addison-Wesley 1995.

[30] Bijnan Bandyopadhyay, Fulwani Deepak, and Kyung-Soo Kim. Sliding Mode Control Using Novel Sliding Surfaces. Springer 2009.